\documentclass{article}
\usepackage[margin=1in]{geometry}
\usepackage{setspace}
\usepackage{graphicx}
\usepackage{color}
\RequirePackage[dvipsnames,usenames]{xcolor}
\usepackage{amsmath}

\begin{document}
  \title{Modeling Social Organizations as Communication Networks}
\author{David Wolpert \\ Santa Fe Institute \\ MIT Aeronautics Dept. \\ Arizona State University \and Justin Grana \\ Santa Fe Institute \and Brendan Tracey \\ Santa Fe Institute   \and Tim Kohler \\Washington State University \and Artemy Kolchinsky \\Santa Fe Institute}
  
\maketitle

\begin{abstract}
We identify the ``organization'' of a human social group as the communication
network(s) within that group. We then introduce three theoretical approaches to
analyzing what determines the structures of human organizations.
All three approaches adopt a group-selection perspective, so that the
group's network structure is (approximately) optimal, given the
information-processing limitations of agents within the social group, and
the exogenous welfare function of the overall group. In the first approach  we use a
new sub-field of telecommunications theory called network coding, and focus
on a welfare function that involves the ability of the organization to convey
information among the agents.  In the second approach we focus
on a scenario where agents within the organization must
allocate their future communication resources when the state of the future
environment is uncertain.  We show how this formulation can
be solved with a linear program.  In the third approach, we introduce an
information synthesis problem in which agents within an organization
receive information from various sources and must decide how to
transform such information and transmit the results
to other agents in the organization.  We propose leveraging the
computational power of neural networks to solve such problems.  These
three approaches  formalize and synthesize work in fields
including anthropology, archaeology, economics and psychology that
deal with organization structure, theory of the firm, span of control
and cognitive limits on communication.  
\end{abstract}

``\emph{Few students of human social dynamics doubt that nations, firms, bands and
other groups are subject to selective pressures . . . group competition may explain
the success of social arrangements.}'' --- [Bowles and Gintis, 2011]

$ $

``\emph{Truly, among Man's innovations, the use of organization to accomplish his ends
is among both his greatest and his earliest. But it is perhaps only in our era, and
even then haltingly, that the rational design of organization has become an object
of inquiry.}'' --- [Arrow, 1964]

$ $

\section{Introduction}

Human social groups are organized in dramatically different ways,
ranging from egalitarian horizontal societies to highly unequal vertical
hierarchies, from groups of decentralized, modular teams to
centralized command and control assemblies. A fundamental problem in
social, behavioral, and economic science is explaining
why particular social groups exhibit their particular organizational structure.
Anthropologists, for instance, are interested in understanding why
post-Pleistocene societies typically demonstrate patterns of
increasing hierarchy and inequality as they grow in size
\cite{evolutionbook}. Economists, likewise, are interested in 
why firms demonstrate differing degrees of
decentralization and hierarchy, and the conditions under which one
form is more advantageous than another
\cite{verthor,primal,radner}. Military theorists seek to understand the relative strengths and
weakness of different command structures, and how those may change within and between campaigns.

Here, we consider organizations that have been subjected to a strong group-selection pressure, for example when
many separate groups have competed with one another for an extended period.
In such scenarios, whatever internal social norms and utility functions lead a social group to adopt a particular organization structure, at the end of the day, those organizations with the best performance best survive. 
Accordingly, to a first approximation we can ignore such considerations.

There are several advantages to focusing on such scenarios. First, it 
allows us to presume that the agents in any thriving organization all
have goals that all closely align with the overall group welfare function. 
This allows us to avoid specifying (or inferring from limited data) variation
in the utility functions of the agents. Similarly, it allows us to
avoid specifying a  bounded rationality game theory solution concept for the collection of
agents. Modeling can thus focus on the
efficiency of the overall organization, and inference from
data can concentrate on determining the organization structure
and the commonly held welfare function. 

There are many factors that have been theorized to drive the form of social organization
in group selection scenarios.
Many researchers have suggested that a fundamental component of 
how well any given social organization performs is how well its members can
collectively gather and process {information} \cite{arrow}.
However much of the work that relates an organization's structure to
informational properties of the organization's agents remains
qualitative and case-specific, even though there is reason to believe
that general and widely applicable theories may exist.  In particular,
in these situations with strong external group pressure, we expect to
mostly see organizations whose communication structure is close to
optimal. These groups are those that are best able to acquire and then
convey information among their members.  


To fill such a fundamental gap in the social sciences, we suggest
three formal and broadly applicable frameworks that analyze how the
informational processing capacities of agents within a group, together
with the group's overall welfare function, determines the group's
optimal organization structure.  Since information processing
abilities is prevalent in the network and communication sciences, we
adopt an ``organizations-as-telecommunication-networks'' paradigm that
explicitly models individuals within an organization as nodes in a
telecommunication network.  This allows us to leverage tools from
those disciplines that have extensive insight into information
processing capabilities.


We begin in section \ref{sec:background} by reviewing the various
ways that researchers across the social sciences have previously attempted to
explain the determinants of social organizations.  In section
\ref{sec:approaches}, we present three theoretical approaches that can
answer various questions relating social organization structure to
information processing capabilities of its agents.  The first approach
leverages a novel sub-field of information theory known as
\emph{network coding}.  To our knowledge, we are the first to employ
the tools of network coding to answer questions in the social sciences.
The details of network coding and its application to social
organization structure are presented in section
\ref{sec:networkcoding}.  Our second approach involves optimizing an
organization's structure when the set of messages it must transmit
through the network is subject to uncertainty.  A main contribution of
our second approach is a {linear program} that solves for the
optimal network structure in such a scenario where the  details  are
given in section \ref{sec:routing}. In our third 
approach, we  seek to model information synthesis among
agents.  In other words, a social organization's success is not just
related to how well it communicates information but how well agents
within the network combine information from various sources and
respond optimally to such information.  
Crucially, the description of each approach is accompanied by a detailed exposition
of the approach's benefits and limitations.  We conclude with section
\ref{sec:synthesis} that highlights how the three approaches can
possibly be synthesized to maximally characterize the properties of
social organizations that arise as a result of information processing
constraints.

We emphasize that
we do not claim that all aspects of social organization can
be explained with the kind of analysis considered here, even of social groups
that have been subject to strong group-level selective pressure. Rather
our goal is to investigate what aspects of social organization can be explained
this way, without the need to imputing more factors. Later work would then
weaken the group selection focus, to incorporate other kinds of
factors that contribute to determining social organization.

\section{Background}
\label{sec:background}
Understanding the structure of social organizations is of interest to
several disciplines including archaeology \cite{arch,population},
management science \cite{managementsci}, sociology \cite{sociology},
political science \cite{polysci} and psychology \cite{psychology}.
It is arguably most prominent in the fields of anthropology and
economics, since in those domains organizations (ranging from small hunter-gatherer tribes to primary
states in archaeology, and from firms to industrial conglomerates in economics) are
often the unit of analysis.  For this reason, much of the previous
work on social organization structure stems from this literature and
we summarize such literatures in turn.

\subsection{Anthropology and Social Organizations}
Early anthropological studies investigating the structure of social
organizations emphasized the importance of kinship and exchange in
organizing small-scale societies \cite{logicprac,andaman}. It is
usually accepted that general forms of social organization€ --- whether
those are firms, societies, or bureaucracies --- €"are correlated to a great
extent with the sizes of the groups. Very large organizations almost
always exhibit hierarchical structures, while very small groups, such
as temporary task groups of fewer than six or so, typically exhibit
``flat'' (non-hierarchical) organizations \cite{contrasting}. Small
bands of shifting membership common among foragers \cite{cores}
are likewise relatively flat in their organization.

Larger and more permanent groups of several hundred people, called
``local groups'' \cite{evolutionbook} or tribes often include
domesticated plants or animals in their diets and their population
density is generally considerably higher than for bands. Enormous
variability in social structure exists in these societies, but three
levels of nested organization beyond the individual are fairly
typical, beginning with families, which reside within kin-based
corporate groups, which in turn may be nested within inter-group
collectivities whose interactions are structured by ceremonial
activities and economic exchange. 

In the broadest sense, most of the anthropological literature focuses
on group size and its relation to organization structure.  More
elaborate models include problems of resource allocation and cultural
norms \cite{norms}. 
However, much of the literature remains case-based and
qualitative.  More fundamentally, while some models assume (implicitly
or explicitly) that information transmission becomes more difficult in
larger groups \cite{j78}, 
the informational transmission is not directly modeled and the direct
causal mechanism that relates information constraints to group size is
obfuscated.  Our theoretical approaches will contribute to this
literature by also explaining how group size partially determines
organization structure.  However, out theoretical approaches expand
beyond the current literature by explicitly modeling the information
transmission constraints a group faces as the group size grows.  We
will then be able to determine what aspects of a social organization
are determined by informational constraints and what aspects are a
result of other factors that arise due to an increasing group size.

\subsection{Economics and the Theory of the Firm}
The economics literature that studies organization structure
falls into one of two categories which together comprise what is known
as ``theory of the firm''.  The first category of models analyzes
organization structure from the standpoint of the 
principal-agent problem and incomplete contracts
\cite{cogame,agency,paprob,innovacost}.  In such models, agents
within an organization---managers and their subordinates within a
firm, for example---have conflicting goals.  The theoretical
predictions then focus on how certain organization structures can
arise such that the principal optimizes the welfare of the
organization while also taking into account the \textbf{different}
incentives that govern the behavior of the other agents.  The second
category of models is related to what is known as ``team theory'' in which
agents in an organization share a similar goal but face coordination
and informational constraints
\cite{verthor,calwel,garsau,williamson}.  Our approach falls into the
second category and specifically focuses on how informational
constraints affect an organization's optimal structure.

Consideration of information's role in determining organization structure dates back at
least to the 1920's and 30's \cite{knight,coase}.  Those early works
suggested that phenomena like informational asymmetries and coordination costs were
the key determinants of an organization's structure.  As an extreme version of this approach, some
argue that without informational limitations, there would be no need
for organizations to act as a coordinating mechanism and each agent
could perform in a way that maximizes welfare by acting independently, without
regard for communication from the other agents
\cite{knight}.  Considering less extreme scenarios, Arrow succinctly
summarized this approach when he wrote ``the desirability of creating
organizations of a scope more limited than the market as a whole, is
partially determined by the characteristics of network information
flows'' \cite{arrow}.  However such claims by Arrow and others were all informal, with no detailed
mathematical model underpinning them;
progress in rigorously establishing (or refuting) the validity of such claims remains
stagnant.  It is our goal to move forward, by building a library of theoretical tools that
allow us to quantitatively determine how the characteristics of
network information flows influence the performance and behavior
of organizations.  

\section{Theoretical Approaches}
\label{sec:approaches}
This section presents our three theoretical approaches.
More of the formal details can be found in the appendix.  

\subsection{Network Coding and Social Organizations}
\label{sec:networkcoding}
Network coding is a relatively new branch of information and
telecommunications theory that began with the seminal work of
\cite{ncmain}.  The main question is straightforward: given a
telecommunications network, what is the most amount of information (in
terms of bits, for example) that can flow from a source to multiple
receivers when links in the network have bandwidth constraints.
Traditionally, this question was addressed when intermediate nodes in
the network were only allowed to ``copy and forward'' information.
The main contribution of network coding is to show that if intermediate
nodes are allowed to perform arbitrary transformations on the
information they receive, the total amount of information transmission
from the sender to the set of receivers can exceed the maximal amount
of transmission when the intermediate nodes are only allowed to copy
and forward.

\begin{figure}
  \centering
  \includegraphics[width=.49\textwidth]{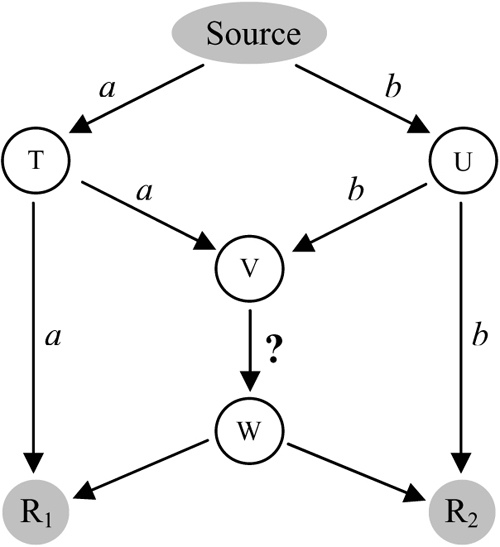}
  \caption{The canonical example of the advantages of network coding. The only way that
    both $R_1$ and $R_2$ can receive both bits $a$ and $b$ is if
    node $V$ transforms its input via the \textbf{XOR}
    operation and then forwards the result of the transformation. If $V$ is
    not allowed to use such network coding, but can only copy and forward its
    inputs, either$R_1$ or $R_2$ will not get both bits.}
\label{fig:butterfly}
\end{figure}

The seminal example of  network coding
is given in figure \ref{fig:butterfly}.  Succinctly stated, the
question is whether nodes $R_1$ and $R_2$ can obtain the value of bits
$a$ and $b$ when each edge in the network can only transfer one bit.
If nodes are only allowed to copy and forward, then the answer to the
question is negative.  For example, if node $V$ forwards the value of
bit $b$ to node $W$, then node $W$ would be able to forward that bit to
$R_1$.   However, it would then be impossible for node $R_2$ to
receive the value of bit $a$ since there are no other channels in
which $R_2$ can receive such a value.  On the other hand, suppose that
the intermediate nodes are able to transform their inputs in an
arbitrary way.  Then, it is possible for both $R_1$ and $R_2$ to
receive the value of both bits $a$ and $b$.  The necessary
transformation would be for $V$ to send the value of the
\textbf{exclusive or} of its inputs $a$ and $b$.  In other words, $V$
would forward $1$ to node $W$  if bits $a$ and $b$ are the same and
and would forward $0$ if bits $a$ and $b$ are different.  Then $W$ would
forward what it receives from $V$ to both $R_1$ and $R_2$.  In this
case, if node $R_1$ receives a bit indicating that the value of $a$ is
$1$ and receives the value $1$ from node $W$, then it can
determine that the value of bit $b$ is $1$. However, if $R_1$ receives
a bit indicating the value of $a$ is $1$ and receives the value $0$
from $W$, then it can determine that the value of node $b$ is $0$.
The same argument can be made for $R_2$ and thus both nodes $R_1$ and
$R_2$ can determine the value of $a$ and $b$.

Drawing on the insight from \cite{ncmain}, network coding has
exploded to be a highly active field.  Not only does it ask whether a
certain amount of throughput is feasible given a network topology and
bandwidth constraints but also asks questions about optimal bandwidth
allocation, algorithmic approaches to creating network codes and
potential benefits of network coding with noisy transmission.  The
main goal of our theoretical approach is to map the questions in
network coding to questions that are relevant in determining the
optimal structure of social organizations.  Table \ref{tab:mapping}
gives an overview of how some of the tools from network coding
can be used to address questions in social organization theory.

\begin{table}
  \centering
  \begin{tabular}{|p{8cm}|p{8cm}|}
    \hline
     Network Coding Question & Social Organization Question \\ \hline
    \vspace{2pt} Given a \textbf{network} and
           \textbf{edge capacities}, what is the most
           amount of information (throughput) that can be transmitted
           from a \textbf{source} to a number of
           \textbf{receivers}? \vspace{2pt}
           & \vspace{2pt} Given an \textbf{organization}
           and \textbf{communication constraints}, what
           is the most amount of information (throughput) that can be
           transmitted from an \textbf{administrator} to
           a number of \textbf{laborers}?  \vspace{2pt} \\ \hline
     \vspace{2pt} How should \textbf{intermediate nodes} transform their inputs such
      that the \textbf{network achieves its maximum   throughput?}\vspace{2pt} &\vspace{2pt}   How should \textbf{middle managers} transform the
      information they receive such  that the \textbf{workers are
        maximally informed}?  \\ \hline
    \vspace{2pt} What is the benefit of adding
           \textbf{extra nodes} in the
           \textbf{network}? \vspace{2pt} & \vspace{2pt}What is the benefit of
           adding \textbf{middle managers} to the
           \textbf{organization}? \vspace{2pt} \\ \hline
  \end{tabular}
  \caption{Typical network coding questions and their possible
    application to social organization theory.} 
  \label{tab:mapping}
\end{table}

We have begun using network coding to determine the benefit of adding
middle managers to a firm, using a minimal model.
In the model, there is an information source and a set of receivers.
The information source can be interpreted either as information from
an external environment or commands  given by an
executive of the firm.  We assume that the organization's welfare is
increasing in the amount of information that the receivers get from
the source.  In other words, the better the organization is at sharing
information, the higher the organization's welfare. However,
communication (in terms of edge capacities) among agents in the
network is costly.  As a crude example, the cost of communication can
represent the opportunity cost of employees in the organization having
a meeting instead of engaging in other productive activities.
Therefore, the goal of the organization is to maximize the amount of
information that the receivers receive from the source minus the cost
of the transmission. This is a standard network utility optimization
problem with associated optimization algorithm that is found in
\cite{ncnum1,ncnum2}. 

\begin{figure}
  \centering
  \includegraphics[width=.49\textwidth]{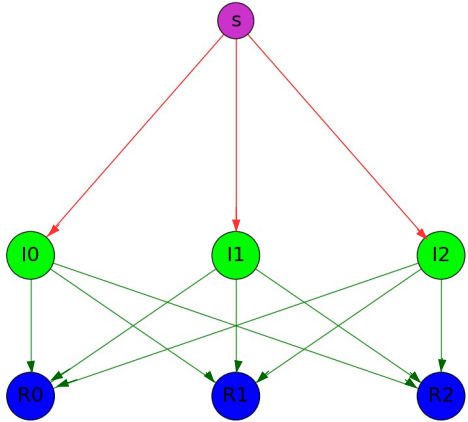}
  \caption{A firm that wants to transmit information from the source
    to the $R$ nodes.}
  \label{fig:noman}
\end{figure}

\begin{figure}
  \centering
  \includegraphics[width=.49\textwidth]{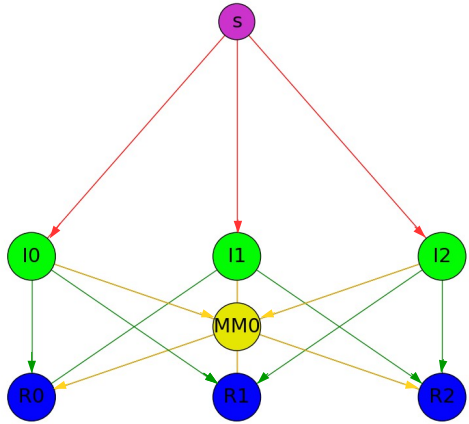}
  \caption{A firm with a middle manager}
  \label{fig:man}
\end{figure}

Figure \ref{fig:noman} depicts a firm without any middle managers.
To add concreteness, we can interpret the source node as an external
environment and the nodes $I0$, $I1$ and $I2$ as observers of the
environment and the links from the $I$ nodes to the $R$ nodes as
communication channels.  The goal is for the $I$ nodes to transmit as much
information about the external environment to the receiver nodes
(denoted $R$) while minimizing the cost of the transmission.  At the
optimum, the $I$ nodes trade off the benefit of relaying information
to the $R$ nodes with the cost of communicating with the $R$ nodes.
While an interesting problem in its own right, we can then extend the
model to include middle managers whose only purpose is to process
information, which is given in figure \ref{fig:man} 
Now the network
optimization problem is repeated and the results show that adding a
middle manager can increase throughput and reduce communication
costs.  More specifically, we can compare the organizations welfare
with and without a middle manager to determine the marginal value of
such a middle manager.  Future work includes extending this scenario to
examine how the benefit of a middle manager changes as  a result of
asymmetries in the network topology and cost parameters.   

\subsubsection{Benefits}
The main benefit of employing network coding to analyze social
organization structure is the trove of untapped analytical and
algorithmic resources that network coding provides.  Previously, many
models of social organizations were overly simplified due to the
difficulty in formulating and solving more complex models.  However,
network coding and its insights provides a theoretical foundation for
modeling complex information transmission problems and
gives several scalable algorithms and analytical techniques that can
be used to solve such problems.  Furthermore, network coding extends
beyond questions regarding pure information transmission.  For
example, network coding can shed light on questions relating to
network robustness \cite{robustness}, resilience \cite{resilience}
and security \cite{security}, all of which may be relevant concerns
for various social organizations.

\subsubsection{Limitations}
There are limitations to using this
network coding approach to analyze social organizations.  The main
limitation is that network coding is mainly concerned with
transferring entire pieces of information from sources to receivers.
However, in real-world organizations it may only be necessary to
transfer \emph{some} information from a source to a receiver.  For
example, a marketing research team does not need to transfer all
possible information it learns to the firm's CEO.  Instead it may
only need to inform the CEO of some high-level insight while at the
same time, the CEO is receiving high level insight from other
departments (finance, accounting, etc.) and must optimally combine
such information.  Technically speaking, the organization does not
need to transfer all information but only \emph{transformations} of
the information.  Due to its disciplinary home in telecommunications,
this problem is not adequately addressed in network coding.  However,
we may be able to use ``intra-session'' coding to establish bounds on
communication rates under more complex message sets \cite{inter}.

\subsection{Contingency Planning}
\label{sec:routing}
Sometimes an organization is uncertain about which communications need
to be made in the future and therefore must allocate communication time
for all possible contingencies.  For example, imagine a large search
and rescue mission in a wildfire scenario. Suppose there is one team
that is dedicated to surveillance and extinguishing  the fire and
another team that is working to locate survivors. The team
that is working to put out the fires must give safety commands to the
search team (i.e. ``don't go east as the fire is spreading that
way'') located in another geographical region.  Depending on
the information the surveillance team receives from the environment,
the surveillance team might have to immediately communicate to the  incident
base instead of communicating to the search team directly.
Therefore, both the search team and the incident base need to be on
``standby'' so that they can hear what the surveillance team might
need to communicate. The search team incurs an opportunity
cost because they cannot proceed to areas where radio contact may be
compromised and the incident base incurs a cost since it must allocate 
attention to the radio in the event that the surveillance team needs
to communicate.  Both the incident base and the search team incur such
a cost \emph{even if the surveillance team does not communicate with
  them.}  In other words, the opportunity cost is incurred as a result
of planning to possibly receive communication, not the communication
itself.  

The above scenario can be modeled as an optimization problem.  More
specifically, the question can be thought of as ``how can agents
allocate sufficient future time such that they are available to
receive all possible communications in any state of the world but
minimize the cost of allocating such time to communication?''  With
certain parametric assumptions, this problem can be modeled as a
\textbf{linear program.}  The solution to the linear program indicates
how each agent (or team) should allocate future time to communicating
with other agents (or teams).  The formal details of the optimization
problem are given in appendix \ref{app:lp}.

This problem can easily be extended to include modifications of the original
scenario.  For example, instead of asking how should agents allocate
resources such that they receive all communications, we can ask how
agents should allocate resources to optimally trade-off the value of
the resources with the benefit of receiving a communication.  This
allows for heterogeneity in the importance of the communication.  Tying this
into the search and rescue scenario, one could imagine that it is more
important for the search team to receive the message ``don't go east
as the fire as spreading that way'' than ``the surveillance team is
undergoing a shift change.''  The linear program framework can easily
be extended to include such a scenario.  

The model can also be extended to include not just costly
communication but constraints on total communication.  For example, it
might be that an agent can only listen to a limited amount of
information in a time period, or can only transmit a limited amount of
information in a time period.  Similarly, agents may be limited in
the number of \emph{distinct} pieces of information they can convey.
Such a constraint can easily be incorporated into the linear program
formulation.  This is an important constraint because it is
well established that individuals face cognitive limits on the number
of interpersonal interactions in a fixed time period \cite{dunbar}.

\subsubsection{Benefits}
There are two main benefits to the contingency planning formulation.  The
first is its tractability and scalability.  In preliminary
experiments, we have solved for the optimal contingency planning in a 
linear program with millions of constraints and variables in less than
1 minute on a single core lap top personal computer.  This indicates that as we expand
the model, computational tractability will be a non-issue.  The second
main benefit of the linear program and contingency planning formulation
is that it allows us to include complicated message requirements.
This is in contrast to network coding in which the same message is
sent from one source to many receivers.  In the contingency planning
formulation, it is possible to include messages from multiple sources
to multiple destinations, where each message contains different
``content.'' 

\subsubsection{Limitations}
Like network coding, the contingency planning approach does not
immediately allow for transformations of information and is only
concerned with full message transmission.  Equally as important, the
contingency planning approach does not include \emph{any} coding at
intermediate nodes and intermediate nodes are only allowed to ``copy
and forward.''  While this may be considered a limitation, there are
indeed a subset of social organizations  where
agents only copy and forward information (prehistoric signaling
networks \cite{signaling} being one plausible example).  Finally, the contingency planning
approach is limited as it does not immediately allow messages to be
corrupted by noise as the message traverses an edge.   

\subsection{Information Synthesis}
\label{sec:neuralnetwork}
The major shortcoming of the network coding and 
contingency planning linear program approaches introduced above is that they do not
model agents that need to synthesize information.  They
are only concerned with the best way to
transfer information within an organization, not how to (have the
agents in the organization) \emph{process} the
information and then use it to take actions back on the
environment.  Furthermore, the network coding  and
contingency planning approaches are limited in the ways in which they
can include noise in the communication channels.

To surmount these issues, we model each agent $i$ as a set of
$k+1$ separate nodes.  Each agent $i$ has a single ``in'' node, represented by $In(i)$,
that provides that agent with the ability to receive information from other
agents.  The remaining $k$ nodes are ``out''
nodes, each representing a distinct ``message'' that the
agent can possibly send. These nodes are represented by $Out(i,j),
j=1...k$.  There is one edge from each $In(i)$  to to each of the associated nodes $Out(i,j),
j=1...k$.  To see how this captures synthesis, imagine the 
following scenario.  An agent receives information from two distinct
sources.  This is represented by two directed edges into $In(i)$.  The
agent can then transform the two pieces of 
information in various ways and send \emph{distinct} messages along
each one of the edges from the agent's ``in'' node to $Out(i,j),
j=1...k$.  There are edges from $Out(i,j), j=1...k$ to other agent's ``in''
nodes which represent communication among agents in a network.
Each node has an associated real number ``value'', where the value
of node $m$ is a function of the values of all of its parents. (Note the similarity
of this model to common neural net architectures.)

This framework allows us to implement many constraints that are
plausible in real world organizations.  For example, we can constrain
the number of edges into $In(i)$, to represent real psychological limits
on how many separate people any given agent can listen to in
a given time period.  The limit on the number of out nodes represents
the similar constraint on how many separate things an agent
can say in a given time period \cite{dunbar}.   Note though that there
is no limit on the total number of other agents that can hear what a given
agent has to say. For example, the CEO of a company may only be able
to say 5 things during a day, but each of those 5 things can be heard
by thousands of others in the organization.

There are several ways to model noise in the
transmission of messages. Perhaps the simplest is to add Gaussian noise to each message that traverses a
channel.  (Note that though for such an approach we should have an upper and lower
bound on the possible values of all nodes, as otherwise the set of all agents
could collectively remove all noise simply by amplifying their signal relative to the
Gaussian standard deviation. )

Some of the agents in the network have inputs from special ``environment''
nodes, which are root nodes. In addition, the values of some agents are
identified as ``actions". So for any given joint value of the
environment nodes, there is a resultant set of values of all the agents,
and in particular a resultant  set of joint actions. This represents a 
single ``wave'', of information from outside of the organization entering
the organization and thereby inducing an action by the organization.
Note the analogy of this model to how the inputs to a neural net induce
states of the hidden nodes that ultimately result in an output of
the neural net. The extension to multiple time-steps, in which the
agents get yet more inputs, while continuing to process old inputs,
is straightforward. (In particular, we can create a time-extended model
of social organizations in much the same way that single-pass neural nets
are extended into recurrent neural nets, to
allow neural nets to process a sequence of multiple successive inputs.)

The welfare of the organization is a function of the joint state of
the environment and the actions. So if there is a distribution over
the state of the environment nodes, there is an associated expected
value of the welfare function. For any given network structure, varying the functions at the agent
nodes will vary that expected welfare. In fact, for any given network
structure, there is an associated maximal value of the expected welfare,
where we maximize over the space of all possible functions at the
agent nodes. Accordingly, by varying the network (subject to
the constraints on input and output nodes of the agents that are described
above), we vary the maximal expected welfare. Accordingly, for any
distribution over environment nodes and welfare function, there
is best-possible network structure, which maximizes the expected welfare.
The group-selection hypothesis is simply that the network structure of a real human social
organization subject to a given welfare function in a given (stochastic)
environment can be well-approximated by this optimal network.

Such a formulation is able to capture a wide range of
characteristics of real world networks. However actually solving for
the optimal network structure (given a welfare function and distribution
over environment states) is a large problem that in general
requires computational optimization.  One possible approach
to solve such an optimization problem would be to formalize
the analogy mentioned above between the overall social organization net
and a neural network. This would allow us to leverage the
computational techniques that have been developed for
training {neural networks}. 
A crucial element of such an approach
would be to use {regularizers} \cite{regular} for ``training the (social organization) neural net''
that capture the cognitive information processing constraints of the agents.  
In particular, we could start with an all-to-all topology and use
an $L_1$ regularizer as a sparsity constraint.  Presumably, this would
push the solution of the network's optimal structure to where many of
the weights are $0$ and therefore can be removed, thereby achieving
(a soft version of) the input-output constraints on the agents mentioned above.


\subsubsection{Benefits}
The main benefit of this ``neural network'' approach is that it provides a
convenient framework for modeling agents that need to synthesize
information.  For example, suppose that one
``message'' among a set of messages is a set of real numbers where one
agent needs to know the average of the numbers and another agent only
needs to know the minimum of the numbers.  An organization modeled in
this neural network fashion would be able to achieve this goal, even if the
communication channels are noisy.  

We can also use this approach to investigate a broad range of social
organization phenomena. How does the (optimal) network structure of an organization
vary as we change the welfare function confronting the organization? How does
it change if we vary the distribution over joint states of the environment? How
robust is an optimal network; if we optimize it for one welfare function and distribution over the
environment, how much does its performance degrade if the welfare function
is changed slightly and/or the distribution over the environment? How
robust is a social organization to internal perturbations, e.g., added noise
on intra-organization transmission lines, the loss of an entire agent, or the like?
If we require that all agents have associated physical locations, and impose
constraints on how far a message can travel in going down an edge, how
does the optimal organization structure change as we improve the communication
technology, so that edges can connect agents who are further apart from
one another?

\subsubsection{Limitations}
A central limitation with this neural network
approach is choosing among the large set of parameterizations of
the functions at the nodes of the agents.
In addition, there
may be several computational issues that arise due to the scope of the
problem.

\section{Conclusion}
\label{sec:synthesis}
We have proposed three theoretical approaches aimed at explaining how
informational capacities impact the structure of a social
organization.  Using such a range of approaches has two main advantages.  The
first advantage is that each approach can address different questions
regarding social organizations.  For example, the network
coding and contingency planning approaches are tractable and
parsimonious but do not allow for complex message demands and
information synthesis.  On the other hand, the synthesis approach
allows for such an analysis.  In  this sense, using all three
theoretical approaches does not confine us to the limitations of one
approach but allows us to optimally ``trade-off'' the benefits and
limitations of each approach in order to fully canvass the properties
of social organization structure.  The second advantage of adopting
three distinct approaches is that it permits a ``robustness check.''
These approaches have enough in common that by pursuing all three,
we can ensure that results from one
theoretical approach do not contradict results from another and
therefore we can be more poised to validate the qualitative
predictions of the general theory.

\section{Acknowledgements}

We recognize the Santa Fe Institute and the NSF for their support under IBSS grant 1620462. 


\bibliographystyle{plain}
\bibliography{ncoprelim.v6}

\appendix

\section{Details of Contingency Planning and Linear Programming Approach}
\label{app:lp}
This section develops a basic problem formulation.

The high level idea is to find the minimum cost network that can
successfully communicate a set of ``broadcasts''. A ``broadcast'' is a
set of ``messages'' that must be communicated simultaneously, and each
``message'' has a single sender and an arbitrary number of
receivers. 


\subsection{The Graph}
The network is a directed (possibly cyclic)  graph, $\mathcal{G} = (S,T,R,
E)$. The nodes in the graph are ``sender`` nodes $s \in S$, ``receiver
nodes`` $t \in T$, and ``relay'' nodes, $r \in R$. The edges in the
graph, $e \in E$, are as follows
\begin{itemize}
\item The pairwise edges from senders to receivers. ($|S| |T|$ total
  edges)
\item The pairwise edges from senders to relayers. ($|S| |R|$ total edges)
\item The pairwise edges from relayers to receivers. ($|R| |T|$ total edges)
\item The pairwise edges between the relayers ($|R|^2$ total edges).
\end{itemize}

Each edge has a weight, $w_e$, representing the possible communication
strength along that edge. Each edge also has a cost, $c_e$, per unit
weight. If $w_e = 0$, the edge cannot be used to communicate, and does
not incur any cost. It is as if the edge does not exist. The edges
with non-zero weight represent the ``existent'' communication network,
i.e. the optimal topology.

\subsection{Messages and Broadcasts}
A ``message'' is an amount of information that must be communicated
from exactly one of the sender nodes to one or more receiver
nodes. Specifically, a message $m$ is a triple, $m = \{s(m), T(m),
\omega_m\}$, where $s(m)$ is the source node $s$ for the message,
$T(m)$ is the set of receiver nodes for the message, and $\omega_m$ is
the amount of information in the message (in the same units as the
edge capacities).

A ``broadcast'', $b \in B$, is a set of messages that must be
communicated simultaneously. Let $M(b)$ be the set of messages indexed
by $b$, and let $m_b$ be an arbitrary message in that set. The network
must have sufficient capacity (sufficient edge weights) to transmit
all messages in a broadcast simultaneously. 

\subsection{Objective}
The objective is to solve the following optimization problem:
\begin{align}
\text{minimize} \quad \sum_e c_e w_e 
\end{align}

\subsection{Constraints}
At a high level, the constraints are that:
\begin{itemize}
\item The edge capacities must be sufficiently high to send each broadcast
\item Within a broadcast, each message must get from all senders to all receivers
\end{itemize}

The constraints will be satisfied using \emph{flow variables}. There
are two kinds of flow variables; flow variables per message, and flow
variables per sender-receiver pair in a message. The per-message flow
variables represent the idea that information can be replicated. A
sender can tell a relayer a piece of information, which can then be
told to multiple receivers. The sender-receiver flow variables
represent the fact that each receiver must receive the entire message.

Let $f_{b,m,e}$ be the flow along edge $e$ for message $m$ in
broadcast $b$. Introduce one such flow variable per possibility (a
total of $ \left( \sum_{b \in B} |M_b| \right) |E|$
variables). Similarly, let $\hat{f}_{b,m,e,t}$ to be the flow
specifically from $s(m)$ to one of $T(m)$.

In the following sub-sections, let $n$ refer to an arbitrary node, let
$p_n$ be the parents to node $n$, and $c_n$ be the children of node
$n$. Similarly, let $e_{p_n}$ be the edges from the parents of $n$ to
$n$, and let $e_{c_n}$ be the edges from $n$ to the children of $n$.

All variables have the additional constraint that they must be
non-negative.

\subsubsection{Message constraints}
Each message must leave the sender and arrive at all of the receiver
nodes for that message. First, we ensure that the entire message is
received by every receiver.
\begin{align}
\sum_{e_{c_s}} \hat{f}_{b,m,e_{c_s},t} = \omega_m \quad &\forall b \in B, ~ m \in M(b), ~ t \in T(m) \\
\sum_{e_{p_t}} \hat{f}_{b,m,e_{p_t},t} = \omega_m \quad &\forall b \in B, ~ m \in M(b), ~ t \in T(m) \\
\sum_{e_{p_n}} \hat{f}_{b,m,e_{p_n},t} = \sum_{e_{c_n}} \hat{f}_{b,m,e_{c_n},t} \quad &\forall b \in B, ~ m \in M(b), ~ t \in T(m) ~ n \in \mathcal{G}_{-\{s,t\}}
\end{align} 
These are the typical flow constraints of goods. The total message
must leave the sender, must get to the receiver, and there must be a
``conservation of message" at each intermediate node. This ensures
that the entire content of each message reaches every receiver.

Secondly, we consider the ``per-message'' flows, which incorporate the
fact that information can be easily replicated. In particular, once a
node receives part (or all) of a message, that same piece of
information can be sent out along any or all of the child edges. This
is distinct from the flow constraints above, since the conservation
constraint becomes an inequality rather than an equality. That is, a
node can only repeat as much of a message as it has heard.
\begin{align}
\sum_{e_{c_s}} f_{b,m,e_{c_s}} = \omega_m \quad &\forall b \in B, ~ m \in M(b) \\
\sum_{e_{p_t}} f_{b,m,e_{p_t}} = \omega_m \quad &\forall b \in B, ~ m \in M(b), ~ t \in T(m) \\
f_{b,m,e_{c_n}} \le \sum_{p_n}  f_{b,m,e_{p_n}} \quad &\forall b \in B, ~ m \in M(b), ~ n \in \mathcal{G}_{-\{s,T(m)\}}
\end{align}
Finally, there is a consistency constraint between these flows --- the sender-recevier flows are upper bounded by the message flows.
\begin{align}
\hat{f}_{b,m,e,t} <= f_{b,m,e} \quad &\forall b \in B, ~ m \in M(b), ~ t \in T(m)
\end{align}

\subsubsection{Satisfying a broadcast}
A single broadcast may contain multiple messages that must be
simultaneously communicated on the network. Each edge must have
sufficient capacity to account for the per-message flow across all the
messages in a single broadcast.
\begin{align}
\sum_{m \in M(b)}  f_{b,m,e} \le w_e \quad \forall b \in B, e \in E
\end{align}
The full linear program is then given by:
\begin{align}
\text{minimize} \quad &\sum_e c_e w_e  \\
\nonumber
\text{subject to\quad}& \\
\nonumber
\sum_{m \in M(b)}  f_{b,m,e} \le w_e \quad &\forall b \in B, e \in E \\
\nonumber
\sum_{e_{c_s}} f_{b,m,e_{c_s}} = \omega_m \quad &\forall b \in B, ~ m \in M(b) \\
\nonumber
\sum_{e_{p_t}} f_{b,m,e_{p_t}} = \omega_m \quad &\forall b \in B, ~ m \in M(b), ~ t \in T(m) \\
\nonumber
f_{b,m,e_{c_n}} \le \sum_{p_n}  f_{b,m,e_{p_n}} \quad &\forall b \in B, ~ m \in M(b), ~ n \in \mathcal{G}_{-\{s,T(m)\}} \\
\nonumber
\hat{f}_{b,m,e,t} <= f_{b,m,e} \quad &\forall b \in B, ~ m \in M(b), ~ t \in T(m) \\
\nonumber
\sum_{e_{c_s}} \hat{f}_{b,m,e_{c_s},t} = \omega_m \quad &\forall b \in B, ~ m \in M(b), ~ t \in T(m) \\
\nonumber
\sum_{e_{p_t}} \hat{f}_{b,m,e_{p_t},t} = \omega_m \quad &\forall b \in B, ~ m \in M(b), ~ t \in T(m) \\
\nonumber
\sum_{e_{p_n}} \hat{f}_{b,m,e_{p_n},t} = \sum_{e_{c_n}} \hat{f}_{b,m,e_{c_n},t} \quad &\forall b \in B, ~ m \in M(b), ~ t \in T(m) ~ n \in \mathcal{G}_{-\{s,t\}}
\end{align}


\end{document}